# Decoherence in dc SQUID phase qubits

Hanhee Paik, S. K. Dutta, R. M. Lewis, T. A. Palomaki, B. K. Cooper, R. C. Ramos,[†] H. Xu,[**]

A. J. Dragt, J. R. Anderson, C. J. Lobb, and F. C. Wellstood

Center for Nanophysics and Advanced Materials, and Joint Quantum Institute

Department of Physics, University of Maryland, College Park MD 20742-4111

**Abstract**

We report measurements of Rabi oscillations and spectroscopic coherence times in an Al/AlO$_x$/Al and three Nb/AlO$_x$/Nb dc SQUID phase qubits. One junction of the SQUID acts as a phase qubit and the other junction acts as a current-controlled nonlinear isolating inductor, allowing us to change the coupling to the current bias leads *in situ* by an order of magnitude. We found that for the Al qubit a spectroscopic coherence time $T_2^*$ varied from 3 to 7 ns and the decay envelope of Rabi oscillations had a time constant $T' = 25$ ns on average at 80 mK. The three Nb devices also showed $T_2^*$ in the range of 4 to 6 ns, but $T'$ was 9 to 15 ns, just about 1/2 the value we found in the Al device. For all the devices, the time constants were roughly independent of the isolation from the bias lines, implying that noise and dissipation from the bias leads were not the principal sources of dephasing and inhomogeneous broadening.

Present Address:       [**] School of Applied Physics, Cornell University, Ithaca NY 14853
              [†] Department of Physics, Drexel University, Philadelphia PA 19104

As the size of a physical system increases beyond the atomic scale, the behavior typically crosses over from the quantum to the classical limit.[1] The scale at which this crossover occurs is not, however, a fundamental one. In principle, a large object that is well isolated from environmental influences can display quantum mechanical nature. Superconducting circuits containing Josephson junctions provide good examples of quantum behavior at larger length scales.[2] Despite the coupling to two-level systems and dielectric layers,[3,4] junctions that are tens of micrometers on a side and which interact with other junctions hundreds of micrometers away have shown superposition of quantum states as well as entangled quantum states.[5-7] Although the junctions were cooled to millikelvin temperatures, it is quite remarkable that in many experiments, the junctions' behavior was monitored by room-temperature amplifiers that were attached through low-pass-filtered meter-length wire leads and still exhibited isolation from room temperature noise. Such large-scale connections to a quantum system from a noisy environment can only be tolerated if the quantum system is sufficiently decoupled from the leads, but not so decoupled that measurements become impossible.

In this paper, we have measured Rabi oscillations and spectroscopic coherence times in order to examine the effects of varying the coupling between a quantum system and noise from the bias leads. Our quantum system is a dc SQUID phase qubit[8] made with Al/AlO$_x$/Al or Nb/AlO$_x$/Nb junctions. In a dc SQUID phase qubit [see Fig. 1(a)], one junction $J1$ is a phase qubit[9] and the rest of the SQUID circuit (fixed inductance $L_1$, isolation junction $J2$ with parallel capacitance $C_2$, and parasitic inductance $L_2$) serves as an on-chip inductive isolation network providing variable isolation from the bias leads.

When the applied flux $\Phi_a$ is held constant, fluctuations $\Delta I$ in the bias current cause small changes $\Delta I_1$ in the current $I_1$ flowing through the qubit junction $J1$. It is useful to define the current noise power isolation factor $r_I \equiv (\Delta I/\Delta I_1)^2$. If the frequency $f$ of the current fluctuations is much less than the resonance frequency of either junction, then

$$r_I = \left(\frac{L_{J1} + L_1 + L_2 + L_{J2}}{L_2 + L_{J2}}\right)^2 \approx \left(\frac{L_1 + L_2 + L_{J2}}{L_2 + L_{J2}}\right)^2 ,$$

(1)

where $L_1 \gg L_{J1}$, as appropriate for our devices. Here

$$L_{J2} = \frac{\Phi_o}{2\pi \sqrt{I_{02}^2 - I_2^2}}$$

(2)

is the Josephson inductance of the isolation junction,[10] $I_{02}$ is the critical current of the isolation junction, $I_2$ is the current going through the isolation junction, and $L_{J1}$ is the Josephson inductance of the qubit junction. We neglect the mutual inductance between the arms of the SQUID.

Equation (1) implies that the current noise power from the bias leads is reduced by a factor of $r_I$ through the inductive isolation network before it reaches the qubit junction and the effective impedance that the bias leads present to the qubit junction is stepped up by a factor of $r_I$. If the current on the leads has a noise power spectral density $S_I(f)$, then the current noise power spectral density $S_{J1}(f)$ that reaches the qubit junction $J1$ is

$$S_{J1}(f) = \frac{S_I(f)}{r_I} \approx S_I(f)\left(\frac{L_2 + L_{J2}}{L_1 + L_2 + L_{J2}}\right)^2 .$$

(3)

We can vary $r_I$ and thus $S_{J1}(f)$ *in situ* because $L_{J2}$ can be changed by varying the current $I_2$ through the isolation junction [see Eq. (2)]. Good isolation can be achieved by choosing $L_1 \gg L_{J2} + L_2$ and the best isolation occurs at $I_2 = 0$, where $L_{J2}$ is a minimum. The choice $L_I \gg L_{JI}$ also ensures that the qubit and isolation junctions are decoupled by the relatively large loop inductance. In this limit the qubit junction behaves like a single Josephson junction qubit.[11] Figure 1(b) shows schematically the corresponding potential energy and energy levels of the 1-D tilted washboard potential experienced by the qubit junction.[12]

A simple characterization of the effects of noise on the qubit is given by measuring the spectroscopic coherence time,

$$T_2^* \equiv \frac{1}{\pi \Delta f} \quad (4)$$

where $\Delta f$ is the full-width in frequency at the half-maximum of the $|0\rangle$ to $|1\rangle$ spectroscopic resonance peak, measured in the low-power limit. Dephasing,[13] inhomogeneous broadening,[14] dissipation,[15] and power broadening[14] all contribute to the spectroscopic width of resonant transitions.[16] Dissipation is the loss of energy by the qubit with timescale $T_1$; dephasing refers to the loss of phase coherence caused by noise at frequencies comparable or faster than $1/T_1$; in contrast, inhomogeneous broadening is due to slow variations in the energy level spacing. Which effect dominates depends on the nature of the measurement and the frequency range of the noise.[16,17]

If $S_{JI}(f)$ is constant below a cut-off frequency $f_c \ll 1/T_1$ and inhomogeneous broadening dominates decoherence, then $T_2^*$ is given by[16]

$$T_2^* = \left[ 1.65 \cdot 2\pi \, \sigma_{JI} \left| \frac{\partial f_{01}}{\partial I_I} \right| \right]^{-1} = r_I^{1/2} \left[ 1.65 \cdot 2\pi \sqrt{S_I(0)f_c} \left| \frac{\partial f_{01}}{\partial I_I} \right| \right]^{-1}. \quad (5)$$

Here $\sigma_{JI} = 2\pi\sqrt{S_{JI}(0)f_c} = 2\pi\sqrt{r_I S_I(0)f_c}$ is the rms current noise in $I_I$ due to $S_{JI}(f)$ and $f_{01}$ is the $|0\rangle$ to $|1\rangle$ transition frequency. Equation (5) shows that $T_2^*$ will depend on $I_I$ because[12]

$$f_{01} \approx f_p \left[1 - \left(\frac{I_1}{I_{01}}\right)^2\right]^{1/4}. \quad (6)$$

From Eq. (6), we find

$$\left|\frac{\partial f_{01}}{\partial I_1}\right| \approx \frac{1}{2} \frac{f_p^4}{f_{01}^3} \left(\frac{I_1}{I_{01}^2}\right), \quad (7)$$

where $f_p = \sqrt{I_{01}/2\pi\Phi_o C_1}$ is the unbiased qubit junction's plasma frequency. We note that for typical biasing conditions $I_1$ is nearly equal to $I_{01}$ and in this limit $\partial f_{01}/\partial I_1$ varies rapidly as a function of $I_1$ or $f_{01}$. From Eqs. (5) and (7), we expect that $T_2^*$ would vary as $f_{01}^3$ and scale with $\sqrt{r_1}$ if $S_{J1}(f)$ is the principal noise source in our qubit.

In contrast, if $S_{J1}(f)$ has a cutoff frequency $f_c \gg 1/T_1$, the effect is to produce dephasing with $T_2^*$ given by[16]

$$\frac{1}{T_2^*} = \frac{1}{2T_1} + \frac{\pi^2 S_I(0)}{r_1}\left(\frac{\partial f_{01}}{\partial I_1}\right)^2, \quad (8)$$

where $T_1$ is the energy relaxation time. In the case of noise dominated decoherence, $T_2^*$ will behave as $r_1 f_{01}^6$, which varies even more rapidly than was the case for Eq. (5).

Additional information about current noise and dissipation from the leads can be obtained by examining Rabi oscillations. The decay time constant $T'$ of the envelope of Rabi oscillations is sensitive to noise at the Rabi frequency, while the shape of the envelope is affected by inhomogeneous broadening from low frequency noise.[17,18] When both dephasing and dissipation are present, the Rabi decay constant $T'$ is related to the energy relaxation time $T_1$ and the coherence time $T_2$ by

$$\frac{1}{T'} = \frac{1}{2T_1} + \frac{1}{2T_2} \tag{9}$$

when there is zero detuning.[19,20,21] Here $T_2$ corresponds to the conventional definition of the coherence time used in NMR and appears in the Bloch equations where $T_2 = 2T_1$ if decoherence is solely due to dissipation.[21]

Since $T'$ and $T_1$ can be measured separately, Eq. (9) allows us to obtain information on $T_2$ even in a system where it is difficult to perform a clean spin-echo measurement. If current noise in the leads is the dominant source of decoherence, we expect $T'$ to scale with $r_I/S_I(f_R)$ where $f_R = \Omega_R/2\pi$ is the Rabi frequency,[17] while if dissipation associated with the lead impedance is the dominant source of decoherence, we expect $T_1$, $T_2 = 2T_1$ and $T' = 4T_1/3$ to scale with $r_I Z_0$, where $Z_0$ is the impedance of the leads at the transition frequency.

To examine the contribution of the leads to decoherence, we measured four dc SQUID phase qubits. Device AL1 [see Fig. 1(c)] was made in our laboratory using photolithography followed by double-angle evaporation of approximately 50 nm thick Al films on an oxidized Si substrate. The 40 μm x 2 μm Al/AlO$_x$/Al qubit junction had a zero-field critical current of $I_{01}$ = 21.28 μA and the device had a single-turn square loop with a 3 μm-width line and sides 300 μm long. Devices NB1, NB2, and NBG were made by Hypres, Inc., from Nb/AlO$_x$/Nb trilayers. Devices NB1 [see Fig. 1(d)] and NB2 had similar layouts consisting of a 6-turn SQUID loop that formed the isolation inductance $L_1$. Device NBG was configured as a gradiometer, with two 6-turn loops in series, wound oppositely to make the device relatively insensitive to uniform external magnetic fields.[22] In device NB1 the trilayer had a nominal critical current density of 100 A/cm$^2$ while for the other two niobium devices the critical current density was nominally 30

A/cm$^2$. For NB1 only, we suppressed the critical currents of the qubit and isolation junctions by applying a small magnetic field in the plane of the junctions so that $I_{01}$ = 34.4 µA for the 10 µm x 10 µm qubit junction; the initial value was $I_{01}$ = 108 µA. The devices were measured in two separate dilution refrigerators using similar detection electronics, microwave filters, and wiring. Devices AL1 and NBG were measured on an Oxford Instruments Kelvinox 25 at a base temperature of 80 mK, while NB1 and NB2 were measured on an Oxford Instruments model 200 at a base temperature of 25 mK. Each refrigerator was enclosed in an rf-shielded room and the devices were shielded against low-frequency magnetic noise by means of a superconducting aluminum sample box and a room-temperature mu-metal cylinder.

For each device, we first measured the current-flux switching characteristics. We found the inductance parameters and rough estimates for the critical current of each junction (see Table I) by fitting the complete $I - \Phi$ curve to the expected characteristics of an asymmetric dc SQUID.[23] We then measured the transition spectrum and Rabi oscillations. For these measurements we simultaneously ramped the current and flux in the appropriate ratio so as to increase the current through the qubit junction linearly with time while keeping the current through the isolation junction fixed.[8,24] Because of the shape of the qubit's washboard potential, higher energy states are more likely to escape via tunneling to the voltage state than lower energy states. As the current through the qubit junction increases, the tilt of the washboard potential increases, decreasing the barrier height and causing the tunneling rates for all levels to increase. We recorded the time at which the switching voltage occurred and from this found the current at which the device tunneled. Repeating this sequence of order $10^5$ times yields a histogram of switching events as a function of current, which we then use to construct a total escape rate $\Gamma$ versus current $I_1$.[25] Since the relatively large loop inductances and critical currents (see Table I) allowed for multiple possible levels of trapped flux in the loop, we used a flux shaking method [24, 26] to initialize the SQUIDs into a desired flux state before each measurement was made.[27]

To vary the isolation from the leads, we used two techniques. For the Nb devices, the measurement was done on each flux state that corresponds to a different but reproducible amount of current circulating in the loop, causing different $I_2$ and $r_I$ values for the same current $I_1$ in the qubit junction. For device AL1, we first used flux shaking to initialize the SQUID into the zero flux state, corresponding to no circulating current in the loop. We next applied a small static offset flux to the SQUID to induce circulating current in the loop, thereby driving current through the isolation junction to set $r_I$.

Figure 2(a) shows the total escape rate $\Gamma$ versus $I_1$ in device AL1 measured at 80 mK for isolation $r_I = 1000$ (solid curve) and $r_I = 200$ (dots). We did not apply microwaves for either of these "background" escape rate curves. The $r_I = 200$ curve shows an overall increase in the escape rate compared to the $r_I = 1000$ curve, as expected if high-frequency noise was present on the bias leads. The broad peaks at 21.02 µA and 21.07 µA in the $r_I = 200$ curve varied in size and location depending on the isolation factor. In separate experiments, we found that these anomalous peaks in the background escape rate were due to noise-induced populations in the $|2\rangle$ and $|3\rangle$ states caused when the $|0\rangle$ to $|2\rangle$ or $|1\rangle$ to $|3\rangle$ transition frequency of the qubit matched the $|0\rangle$ to $|1\rangle$ resonance frequency of the isolation junction.[11,28,29] We note that the total escape rate is given by $\Gamma = \sum_i \rho_i \Gamma_i$ where $\rho_i$ and $\Gamma_i$ are the normalized occupation probability and escape rate from level $i$. Since the escape rates increase by two to three orders of magnitude for each successive level, very small populations in the excited states are detectable. For example, in Fig. 2(a) at $I_1 = 21.07$ µA we estimate that the probability of occupying $|2\rangle$ increases by only about 50 parts per million when $r_1$ changes from 1000 to 200.

We next measured the total escape rate $\Gamma$ versus $I_1$ while applying microwaves for $r_1 = 1000$ [see Fig. 2(b)]. For all the devices, clear resonant peaks were found in the 6-8 GHz range and the dependence of the $|0\rangle$ to $|1\rangle$ transition on current was in good agreement with that expected for a single Josephson junction. We fit the data to the expected spectrum of a single Josephson junction to determine the qubit critical current and capacitance (see Table I). With the power set low enough that power broadening was not apparent, we measured $\Delta f$ for the $|0\rangle$ to $|1\rangle$ transition and then applied Eq. (4) to obtain $T_2^*$.

Figure 3(a) shows $T_2^*$ versus $r_1$ for device AL1 measured at 80 mK and 7.45 GHz. We find that $T_2^*$ varies between 3 ns and 7 ns in an apparently random fashion as $r_1$ varies by an order of magnitude. We note that $T_2^*$ showed neither the linear, nor the square root dependence on $r_1$, predicted by Eqs. (5) and (8). For comparison, Fig. 3(b) shows $T_2^*$ versus $r_1$ for device NB1 measured at 7.2, 7.3, 7.4, and 7.5 GHz at 25 mK. It shows an apparent random variation between about 3 and 6 ns. Devices NB2 and NBG were measured at fixed isolations of $r_1 = 2500$ and $r_1 = 2300$, respectively, and showed $T_2^*$ values that were about the same as for NB1 (see Table I). Furthermore, none of the four devices showed the strong systematic dependence of $T_2^*$ on $I_1$ or $f_{01}$ predicted by Eqs. (5) or (8) [see Fig. 3(b), for example].[22]

Finally, we measured Rabi oscillations in the four devices by applying microwave current at the $|0\rangle$ to $|1\rangle$ transition frequency and then monitoring $\Gamma$ as a function of time from when the microwaves started. Figure 4 shows Rabi oscillations in the escape rate for device AL1 at 80

mK with a 7 GHz drive for $r_I = 1000$ [Fig. 4(a)] and for $r_I = 200$ [Fig. 4(b)]. The solid curves are $\chi^2$–fits to our phenomenological model for the oscillations

$$\Gamma = g_0 + g_1\{1 - e^{-(t-t_0)/T'}\cos[\Omega_R(t-t_0)]\} + g_2(1 - e^{-(t-t_0)/T_0}). \tag{10}$$

The fitting parameter $T'$ gives the decay time constant of Rabi oscillations, $g_1$ sets the amplitude of the oscillations, and $t_0$ is the time when the microwave power was turned on. The parameter $T_0$ and the $g_2(1 - e^{-(t-t_0)/T_0})$ term account for the rise time of the microwave pulse, and emulate the effect of increased occupancy in higher levels such as $|2\rangle$ at high microwave power. The parameter $g_0$ combined with $g_2(1 - e^{-(t-t_0)/T_0})$ accounts for the initial background escape rate, and includes contributions from $|0\rangle$, $|1\rangle$, and higher levels caused by noise or thermal excitation. We found that including $g_0$ and $g_2$ significantly improved the quality of the fits but had little effect on the estimated Rabi decay time $T'$.

Comparing Figs. 4(a) and 4(b), we note the increased escape rate at $t = 0$ for $r_I = 200$, as expected from an increase in noise-induced transitions to an excited state due to a decrease in the isolation. However, fitting Eq. (10) to the data yields $T' = 33$ ns for the $r_I = 200$ curve and $T' = 28.2$ ns for the $r_I = 1000$ curve. We also measured Rabi oscillations for $r_I = 1000$ at six different Rabi frequencies (from 111 MHz to 188 MHz) and for $r_I = 200$ at twelve different Rabi frequencies (from 33 MHz to 122 MHz). The range of $T'$ was 20 ns to 28 ns for $r_I = 1000$ and 20 ns to 33 ns for $r_I = 200$. Thus the decay time constant of Rabi oscillations did not scale with $r_I$ or $\sqrt{r_I}$.

If the decoherence were entirely due to dissipation, we would expect the coherence time $T_2 = 2T_1$, and the Rabi decay time $T' = 4T_1/3$. To test whether our qubits are dissipation limited, we found $T_1$ by measuring a series of background escape rates at temperatures from 80 mK to 200 mK in the maximally isolated case ($r_I = 1000$). We noted the size and location of the shoulder of each thermal escape rate to estimate $T_1$.[30] In device AL1, this procedure yielded $T_1 \approx 20$ ns, a value for which $T' = 4T_1/3 \approx 27$ ns.

We also observed that $T'$ was independent of the isolation factor $r_I$ in the Nb devices. For example, Fig. 5 shows Rabi oscillations in NB1 at 25 mK with a 7.6 GHz microwave drive for $r_I = 1300$ [Fig. 5(a)] and $r_I = 450$ [Fig. 5(b)]. Fitting to Eq. (10) yields $T' \approx 12$ ns for $r_I = 1300$ and $T' \approx 15$ ns for $r_I = 450$. We found $T'$ from 9 to 15 ns for $r_I$ in the range of 50 to 1300. As with device AL1, no apparent systematic dependence of $T'$ on the isolation was shown in NB1.

However, there was one significant difference in behavior between NB1 and AL1. From $T_1$ measurements, we found $T_1 \approx 14$ ns for NB1. If the decoherence were entirely due to dissipation, we would expect $T' = 4T_1/3 \approx 19$ ns, which manifestly disagrees with the observed $T'$ data for this device. Other Nb devices such as NB2 and NBG showed qualitative behavior and quantitative results that were very similar to device NB1 (see Table I). This disagreement means that an additional dephasing mechanism is present beyond that due to dissipation.

Table I summarizes the parameters and main results for all four devices. The fact that $T'$ and $T_2^*$ did not depend systematically on the isolation from the leads implies that neither current noise from the leads nor dissipation in the leads is the main source of decoherence in these devices, even though we observe clear noise-induced transitions in the escape rate that vary with the isolation. We also note that $T_1$ was in the 15 to 20 ns range for all four devices, but the aluminum qubit AL1 showed a substantially longer Rabi decay time $T'$ than the Nb devices.

Another possible source of decoherence is local 1/f flux noise of unknown origin that has been found in other SQUIDs at millikelvin temperatures.[31,32] Decoherence from such a source would be largely independent of $r_I$ but would depend on $\partial f_{01}/\partial I_1$, which in turn depends strongly on $f_{01}$ or $I_1$. Our data do not support such a dependence. Also, $T'$ and $T_2^*$ for the gradiometer NBG were very comparable to those for magnetometers NB1 and NB2, but shorter than for magnetometer AL1.[22] This strongly suggests that spatially uniform flux noise was not responsible for the short coherence times in our dc SQUID phase qubits.

Simmonds *et al.* and Martinis *et al.*[3,4] have pointed out that the likely source of decoherence in phase qubits is spurious two-level charge fluctuators that reside in the substrate or dielectric layers. The fact that $T'$ for device AL1 was two times longer than for the three Nb devices is suggestive of a materials related effect. AL1 had a thermally grown $AlO_x$ tunnel barrier, native oxide on the exposed metal surfaces, and the thermally grown $SiO_2$ layer on a Si substrate but no wiring insulation layer. In contrast, the Nb devices had all of the above plus sputtered $SiO_2$ insulation layers between the wiring layers. While we have not seen clear spurious resonant splittings (down to a resolution of about 10 MHz) in spectroscopic data on AL1, we have identified small apparent splittings of about 5-10 MHz in NB1.[33]

In conclusion, we have measured the spectroscopic coherence time $T_2^*$ and the time constant $T'$ for the decay of Rabi oscillations in four dc SQUID phase qubits with variable coupling to the leads. From these measurements we can determine the impact of the leads on noise and decoherence in the dc SQUID phase qubits. We found that varying the isolation from the leads produced no systematic effect on either $T_2^*$ or $T'$, and that with comparable isolation, the aluminum device had a coherence time $T_2$ that was two to three times longer than that of the Nb devices. This implies that the leads are not the dominant source for decoherence in these devices. Instead, our data are consistent with a local, materials related source of decoherence.

This work was funded by the NSF through the QuBIC Program (grant number: EIA0323261), the NSA through the Laboratory for Physical Sciences, and the state of Maryland through the CNAM, formerly the Center for Superconductivity Research. We acknowledge many helpful conversations with A. J. Przybysz, H. Kwon, K. Mitra, F. W. Strauch, P. R. Johnson, B. Palmer, M. Manheimer, M. Mandelberg, J. M. Martinis and R. W. Simmonds.

Table I. Parameters for dc SQUID phase qubits AL1, NB1, NB2, and NBG. $I_{01}$, $L_{J1}(0)$, and $C_1$ are critical current, zero-bias Josephson inductance, and capacitance of the qubit junction, respectively, while $I_{02}$, $L_{J2}(0)$, and $C_2$ are corresponding values for the isolation junction. $T$ is the temperature and "Range of $r_I$" gives range of isolation factors $r_I$ examined for $T_2^*$ and Rabi measurements. $T_2$ is the coherence time found from $T_1$, $T'$, and Eq. (9). The last row indicates that only for device AL1 can dissipation account for all the decoherence.

| Device | AL1 | NB1 | NB2 | NBG |
|---|---|---|---|---|
| $I_{01}$ (μA) | 21.28 | 34.3 | 19.4 | 23.0 |
| $I_{02}$ (μA) | 9.45 | 4.4 | 7.5 | 3.8 |
| $L_{J1}(0)$ (pH) | 13.2 | 9.6 | 17.0 | 13.9 |
| $L_{J2}(0)$ (pH) | 44.5 | 75 | 43.9 | 84.9 |
| $L_1$ (nH) | 1.24 | 3.52 | 3.39 | 4.54 |
| $L_2$ (pH) | 5 | 25 | 25 | 12 |
| $C_1$ (pF) | 4.1 | 4.4 | 4.4 | 4.1 |
| $C_2$ (pF) | 2.1 | 2.2 | 2.2 | 2.0 |
| T (mK) | 80 | 25 | 25 | 80 |
| Range of $r_I$ | $50 < r_I < 1000$ | $50 < r_I < 1300$ | $500 < r_I < 2500$ | 2300 |
| T' (ns) | 20 - 33 | 9 - 15 | 11 - 15 | 9 - 15 |
| $T_2^*$ (ns) | 3 - 7 | 3 - 6 | 3 - 8 | 4 - 8 |
| $T_1$ (ns) | ~ 20 | ~ 14 | ~ 17 | ~ 15 |
| Estimated $T_2$ (ns) | ~ 40 | 7 - 16 | 8 - 13 | 6 - 15 |
| $T_2 \approx 2T_1$ ? | Yes | No | No | No |

**Figure Captions**

FIG. 1. (a) Schematic of dc SQUID phase qubit. Current bias I and flux bias $\Phi_a = MI_\Phi$ are used to control currents $I_1$ through qubit junction $J1$ and $I_2$ through isolation junction $J2$. $C_1$ and $C_2$ are capacitances of the qubit and the isolation junctions, respectively. (b) 1-D tilted washboard potential. $\Gamma_0$, $\Gamma_1$, and $\Gamma_2$ are the escape rates from the ground state, the first excited state and the second excited state. (c) Photograph of SQUID AL1. $J1$ (left) has unused contact pads attached for adding auxiliary coupling components. (d) Photograph of dc SQUID phase qubit NB1.

FIG. 2. (a) Background escape rates $\Gamma$ versus current $I_1$ for device AL1 for $r_I = 1000$ (solid curve) and $r_I = 200$ (dots). Escape rates were measured at 80 mK. (b) Escape rates $\Gamma$ versus $I_1$ with (dotted) and without (solid) application of 7 GHz microwaves for AL1 for $r_I = 1000$ when the device parameters were different.

FIG. 3. (a) Spectroscopic coherence time $T_2^*$ versus isolation factor $r_I$ for device AL1 measured at 7.45 GHz and 80 mK. (b) $T_2^*$ versus $r_I$ for SQUID NB1 for microwave frequencies of 7.2 (crosses), 7.3 (open circles), 7.4 (closed circles) and 7.5 GHz (open squares) at 20 mK.

FIG. 4. Rabi oscillations in the escape rate of device AL1 at 80 mK generated by turning on a 7 GHz microwave source at time t = 0 at $I_1 = 21.07$ μA. Crosses show Rabi oscillation data and solid curves show the best fit to Eq. (10) for (a) $r_I = 1000$ (best isolation) and (b) $r_I = 200$ (poor isolation). For $r_I = 1000$ the fitting parameters were $t_0 = 1.8$ ns, $g_0 = 2.8$/μs, $g_1 = 4.9$/μs, $g_2 = $

3.1/μs, $T' = 28.2$ ns, $T_0 = 1.7$ ns and $\Omega_R/2\pi = 111$ MHz, while for $r_I = 200$ the fitting parameters were $t_0 = 2.5$ ns, $g_0 = 4.3$/μs, $g_1 = 2.7$/μs, $g_2 = 4.5$/μs, $T' = 33.0$ ns, $T_0 = 3.8$ ns and $\Omega_R/2\pi = 97$ MHz. The larger $g_0$ for the Rabi oscillation data with $r_I = 200$ implies that noise-induced transitions are present in the background escape rate.

FIG. 5. Rabi oscillations in device NB1 at 25 mK generated by switching on a 7.6 GHz microwave source at time t = 0 at $I_1 = 34.07$ μA. Crosses show data and solid curves show fits to Eq. (10) for: (a) $r_I = 1300$ (best isolation) and (b) $r_I = 450$ (poor isolation). For $r_I = 1300$ the fitting parameters were $t_0 = 1.3$ ns, $g_0 = 5.1$/μs, $g_1 = 22$/μs, $g_2 = 6.9$/μs, $T' = 12$ ns, $T_0 = 0.9$ ns and $\Omega_R/2\pi = 172$ MHz, while for $r_I = 450$ the fitting parameters were $t_0 = 1.4$ ns, $g_0 = 5.5$/μs, $g_1 = 22$/μs, $g_2 = 7.1$/μs, $T' = 15$ ns, $T_0 = 0.9$ ns and $\Omega_R/2\pi = 169$ MHz.

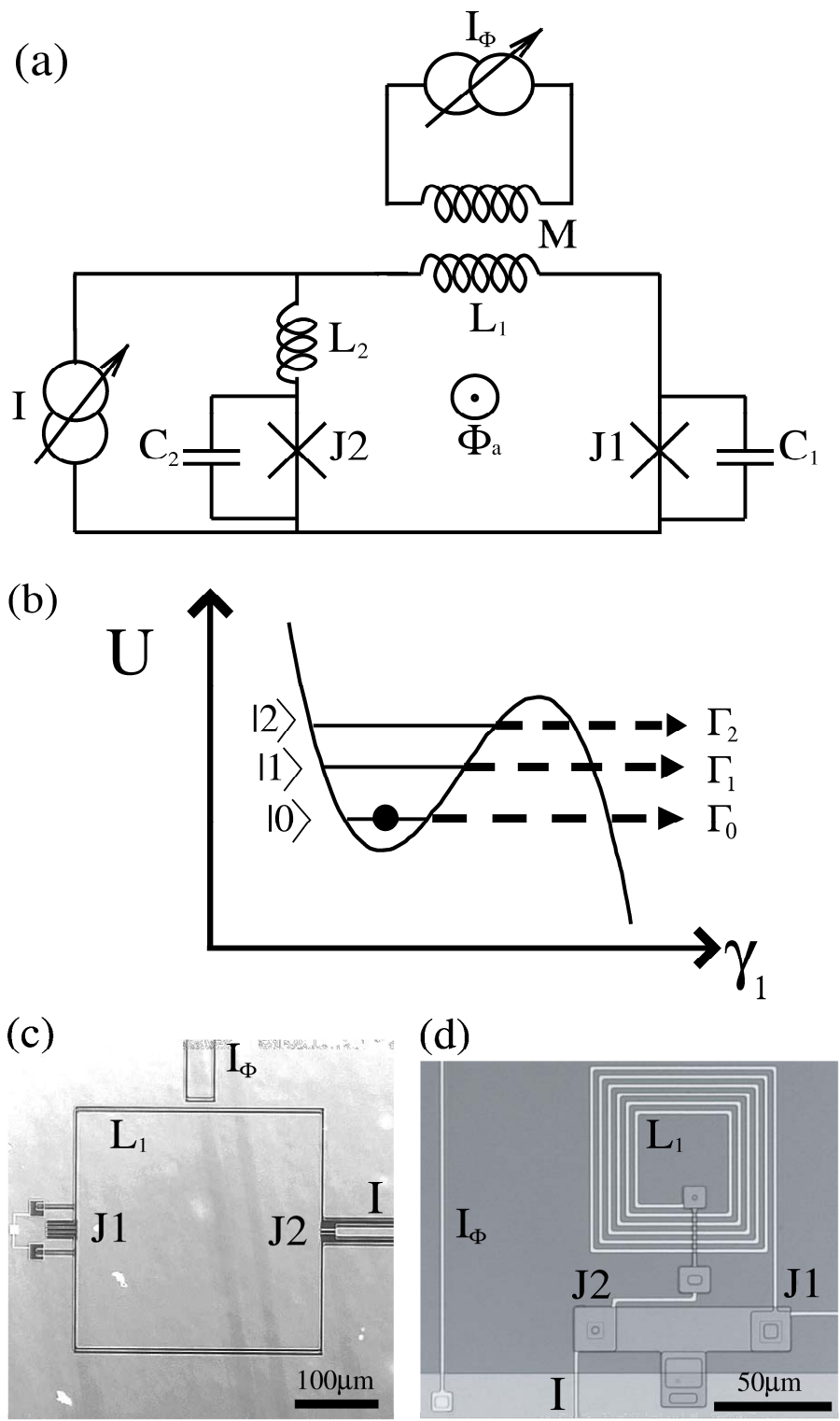

FIG.1, Paik et al.

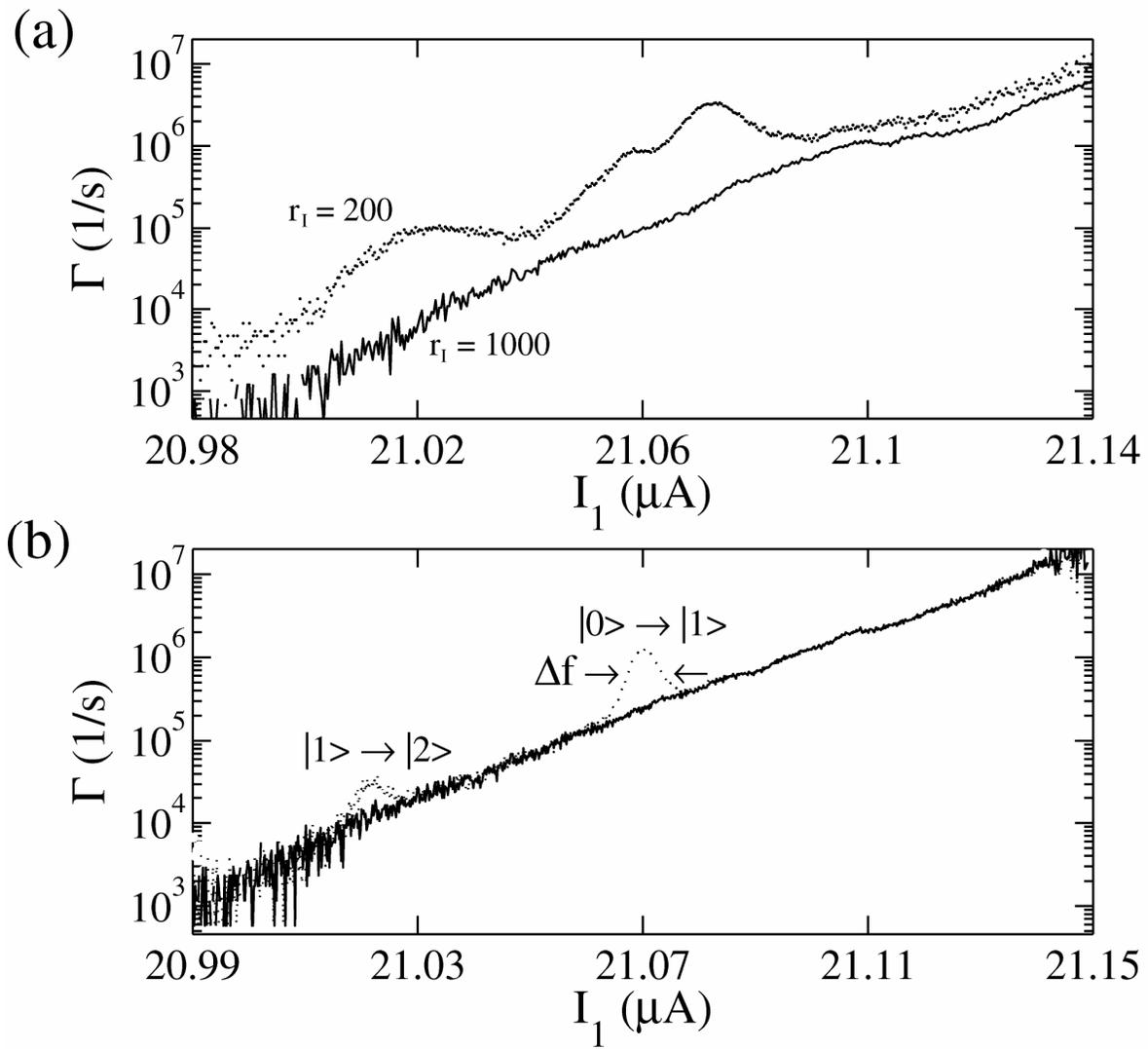

FIG. 2, Paik *et al.*

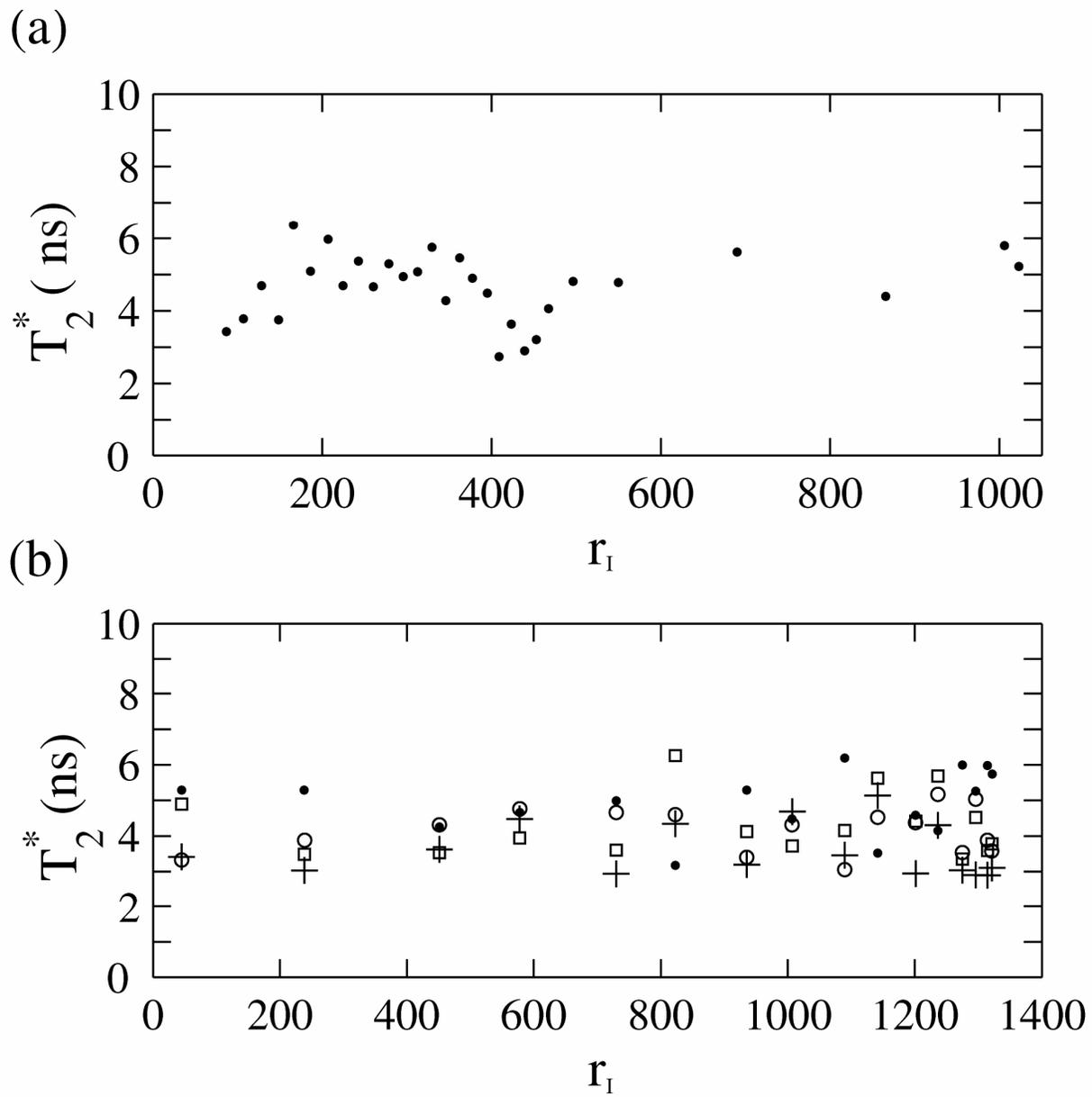

FIG. 3, Paik *et al.*

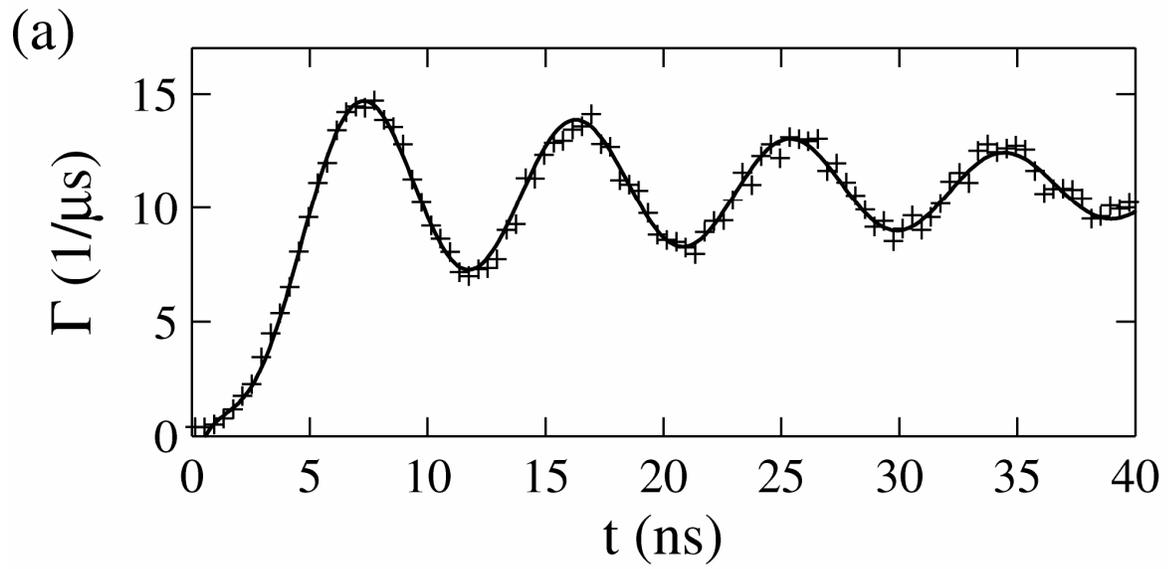

(a)

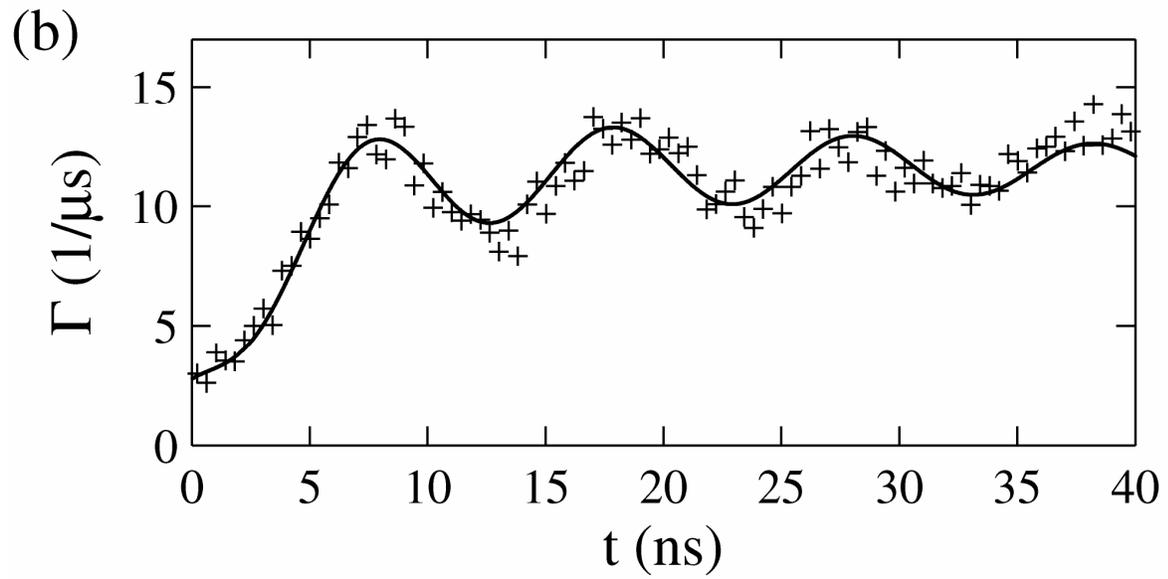

(b)

FIG. 4, Paik *et al.*

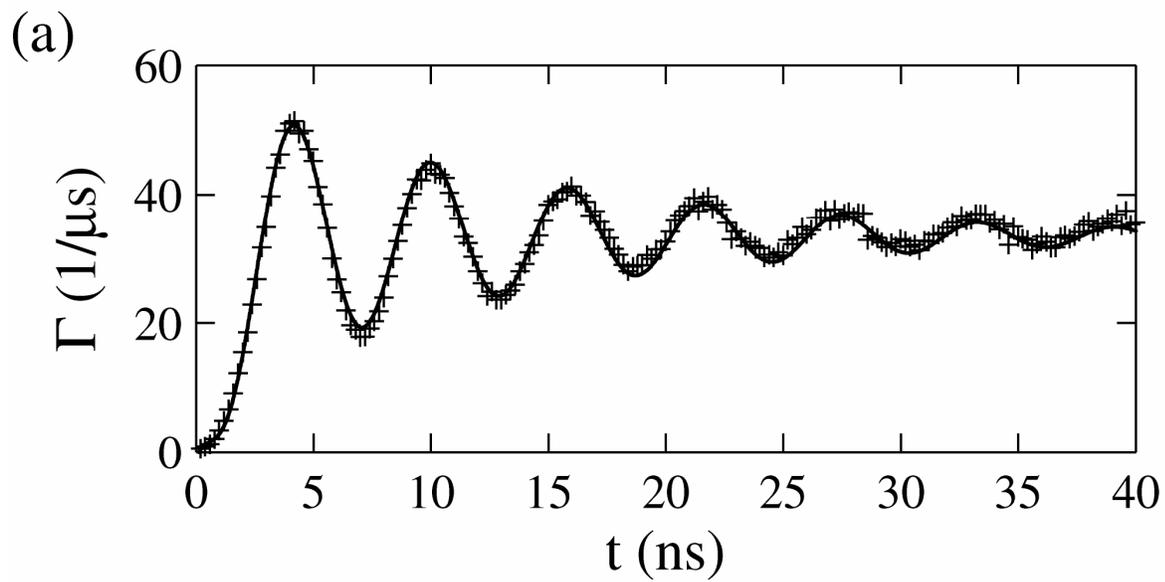

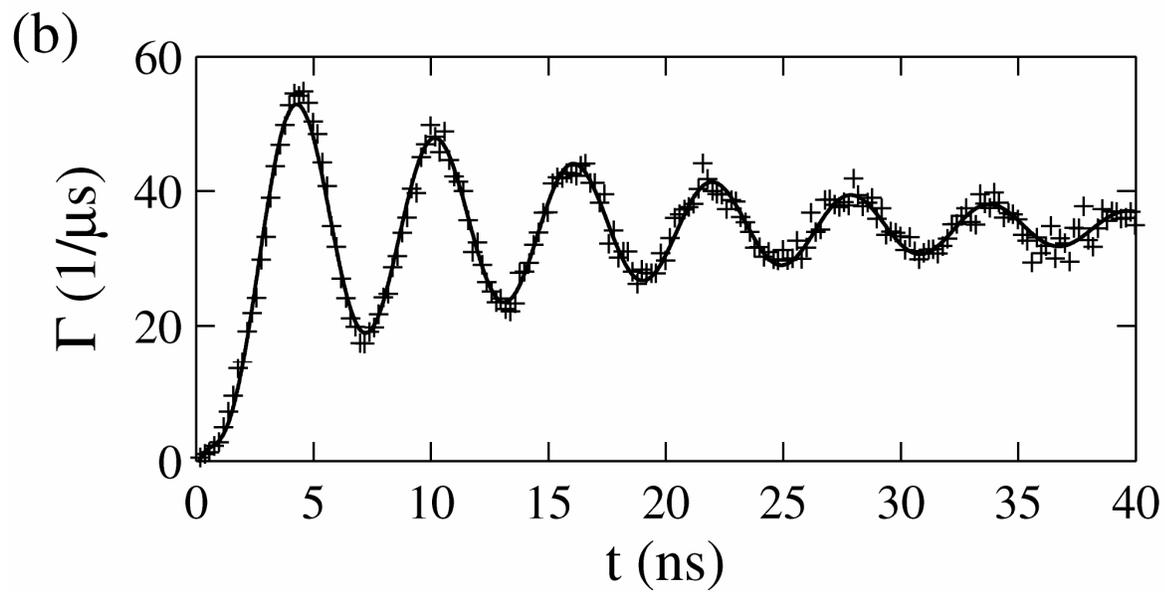

FIG. 5, Paik *et al.*